\documentclass[journal=jacsat,manuscript=article]{achemso}

\usepackage[version=3]{mhchem} 
\usepackage{textcomp}
\usepackage{gensymb}
\usepackage[export]{adjustbox}

\newlength{\figurewidth}
\newlength{\figurewidthdouble}
\author{Jie Zhan}
\affiliation{Department of Chemistry, University of Georgia, Athens, GA, 30601}
\author{Nicholas D. Cooper}
\affiliation{Department of Chemistry, University of Georgia, Athens, GA, 30601}
\author{Melanie A. R. Reber}
\affiliation{Department of Chemistry, University of Georgia, Athens, GA, 30601}
\email{mreber@uga.edu}

\title
  {Three-Wave Mixing between Continuous-Wave and Ultrafast Lasers}

\abbreviations{IR,UV}

\begin{document}
\begin{center}\date{May 07, 2024}\end{center}

\begin{abstract}
  Ultrafast optical frequency combs allow for both high spectral and temporal resolution in molecular spectroscopy and have become a powerful tool in many areas of chemistry and physics. Ultrafast lasers and frequency combs generated from ultrafast mode-locked lasers often need to be converted to other wavelengths. Commonly used wavelength conversions are optical parametric oscillators, which require an external optical cavity, and supercontinuum generation combined with optical parametric amplifiers. Whether commercial or home-built, these systems are complex and costly.  Here we propose an alternative, simple, and easy-to-implement approach to tunable frequency comb ultrafast lasers enabled by new continuous-wave laser technology. Sum-frequency generation between a Nd:YAG continuous-wave laser and a Yb:fiber femtosecond frequency comb in a beta-barium borate (BBO) crystal is explored. The resulting sum-frequency beam is a pulsed frequency comb with the same repetition rate as the Yb:fiber source. SNLO simulation software was used to simulate the results and provide benchmarks for designing future system to achieve wavelength conversion and tunability in difficult spectral regions.
  
\end{abstract}

\section{Introduction}
Combining high temporal resolution and wide spectral coverage, ultrafast optical spectroscopy is used in a wide range of fields including chemical dynamics\cite{nibbering_AnnRev2005,Butler_CCR2007}, photo-induced dynamic processes {\cite{maiuri2019ultrafast,Didonato_BiochBiophys2015}}, and materials characterization\cite{Munson_JMCC2019}. Ultrafast lasers are also used to generate optical frequency combs, the narrow spectral linewidths provide ultra-high precision in both time and spectral resolution {\cite{cingoz2012direct,udem2002optical,picque2019frequency}}. 
Frequency combs have found application in precision metrology and molecular spectroscopy \cite{cingoz2012direct,keilmann2012mid,bartels200910,li2016high}. Mode-locked lasers, including Ti:Sapphire lasers and fiber lasers, are the most widely-used sources for ultrafast pulse generation, including frequency comb lasers. To cover the spectral regions of interest to molecular spectroscopy and other applications, it is generally necessary to frequency convert the light external to the laser. Common setups for creating tunable radiation from a mode-locked ultrafast laser is an optical parametric oscillators (OPO) or optical parametric amplifiers (OPA) combined with supercontinuum generation. 

Ultrafast OPOs have been built for a range of wavelength regions, pulse durations, and tunability bandwidths. They all utilize an optical cavity with an intracavity nonlinear crystal and are pumped with mode-locked laser.\cite{geesmann2023rapidly,mevert2021widely,deckert2023sub} OPOs involve coupling the incoming laser to an external optical cavity, which contributes to the complexity and environmental demands of an instrument. Active feedback and control are necessary for high performance OPOs and involved PID controllers and feedback mechanisms. \cite{reid2010advances,kobayashi2015femtosecond} Frequency tunability of ultrafast frequency comb lasers also use OPOs pumped by mode-locked lasers\cite{kobayashi2015femtosecond,chen2019tunable}.
 
Ultrafast OPAs use a high-power pulsed laser to amplifier a seed laser in a nonlinear crystal. The seed is often a tunable source made through supercontinuum generation.\cite{dalla2023mid,xu2020high} OPAs are commonly seeded with kHz repetition rate lasers, which can undergo supercontinuum generation in a range of materials, such as sapphire and calcium fluoride. MHz repetition rate lasers have lower peak powers for the same average power so supercontinuum generate is usually done in a nonlinear fiber, which can have stability problems. This limitation on supercontinuum materials ultimately presents a limitation on using them for high-repetition rate tunable lasers, even in an OPA configuration. Additionally, OPAs require the overlap in both space and time of the pump laser and the seed laser, which adds to the experimental complexity.\cite{brida2009few,manzoni2016design}
 
To get around the challenges of OPOs, OPAs, and supercontinuum generation, and to take advantage of the developments in stable, tunable, continuous wave (CW) sources coming on the market, we revisit the possibility of using difference frequency generation (DFG) and sum frequency generation (SFG) to make tunable ultrafast light without an OPO configuration. We investigate three-wave mixing between CW and ultrafast laser in a nonlinear crystal and model it using theory, to lay the foundation for further implementation for a range of wavelength regions. 
 
Specifically, three-wave mixing between a CW and an ultrafast frequency comb is achieved through sum frequency generation in a BBO crystal between a ultrafast Yb:fiber frequency comb centered at 1055 nm and an Nd:YAG CW laser centered at 1064 nm. The ultrafast Yb:fiber laser is an optical frequency comb.\cite{reber2016cavity,NickThesis} Each optical comb tooth has a frequency, $\nu_n(Yb:fiber)$, and will undergo sum-frequency generation with the single-frequency Nd:YAG laser: 
\begin{equation} \label{eq:1}
\nu_{m(SFG)} = \nu_{n(Yb:fiber)} + \nu_{(YAG)}   
\end{equation}
The resulting SFG light will be a frequency comb with the frequency of each comb tooth, $\nu_{m(SFG)}$, equal to the sum frequency of a single Yb:fiber comb tooth with the CW laser. This is analogous to that seen in difference frequency generation between two frequency combs{\cite{schliesser2012mid}}. 
 
To the best of our knowledge, three-wave mixing between CW and pulsed laser has been reported only once before by Salhi et. al{\cite{SALHI2007198}}, who demonstrated SFG in a beta barium borate (BBO) crystal between a femtosecond Ti:Sapphire laser with 130 fs pulse duration, center wavelength of 797 nm, and a GaInAsP/InP CW laser, with emission wavelength at 1551 nm. Their resulting SFG beam had a center wavelength of 527 nm, a power of 10 pW, and conversion efficiency of $7 \times 10^{-10}$. A systematic study, optimization of conversion efficiency, and comparison to computational predictions is still required to be able to use this method for a tunable laser design. Here we expand upon that first observation by systematically characterizing the SFG between a CW laser and ultrafast frequency comb laser and comparing the results to calculations made with SNLO, a software to perform detailed simulations of nonlinear mixing processes in nonlinear crystals{\cite{snlo}}. 

 \begin{figure}[ht]
\centering
\fbox{\includegraphics[width=\figurewidthdouble]{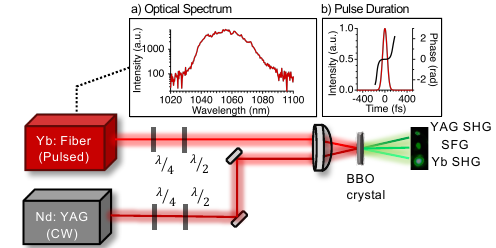}}
\caption{Experimental Setup Scheme with a photo of the YAG SHG, Yb SHG, and SFG beam. Inset: a) Spectrum of Yb:fiber comb and b) Pulse profile from FROG measurement.}
\label{fig:fig1}
\end{figure}

\section{Experimental Setup}

The experimental setup is shown in Figure \ref{fig:fig1}. The ultrafast laser is a homebuilt 85 MHz Yb:fiber laser frequency comb and amplifier{\cite{YbLaserSystem,NickThesis}} with a center wavelength of 1055 nm, 20 nm FWHM (Figure  \ref{fig:fig1}a), and a maximum power of 9 W. An intracavity bulk EOM and $\mu$m-scale piezo-electric adjustments stabilize the comb. It has a 97 fs pulse duration (Figure \ref{fig:fig1}b) measured by the frequency-resolved optical gating technique (Swamp Optics GRENOUILLE). In this work, comb powers from 200 to 500 mW are used. The CW source is a stable non-planar ring oscillator Nd:YAG laser (Coherent Mephisto) with a center wavelength of 1064 nm, 3 kHz linewidth, and up to 500 mW output power, of which 20 to 110 mW is used. The beam sizes of the two sources are made to match by Keplarian telescopes, waveplates to align the polarization of both beams, and then both are focused by a 50 mm plano-convex lens to a 5 x 5 x 1.0 mm BBO crystal with a Type I phase matching scheme. The ABCD matrix method is used to calculate the beam size and radius of curvature of both lasers at the BBO crystal, based upon beam profile measurements made with the knife-edge method.

Sum-frequency generation between the CW laser and the frequency comb source is achieved under a non-collinear regime, with a relative angle of 11\degree\ and a 5.5\degree\ angle to the crystal normal. This is analogous to an autocorrelation geometry where the resulting CW (YAG) SHG, pulsed (Yb) SHG, and the SFG beam are spatially separated. A photograph of the three generated beams is shown in Figure \ref{fig:fig1}. The incident angle of the crystal can be tuned to optimize the SHG of either the YAG or Yb lasers or the SFG between the two. The image shows an incident angle where all three were visible with the camera. Generated beam power was optimized prior to taking data. The SFG and SHG beams are characterized with a grating spectrometer (ASEQ LR 1), a power meter (Thorlabs S130C), and an RF spectrum analyzer (Rigol DSA 815).

\section{Results and Discussions}

\begin{figure}[htbp!]
\centering
\fbox{\includegraphics[width=0.7\figurewidthdouble]{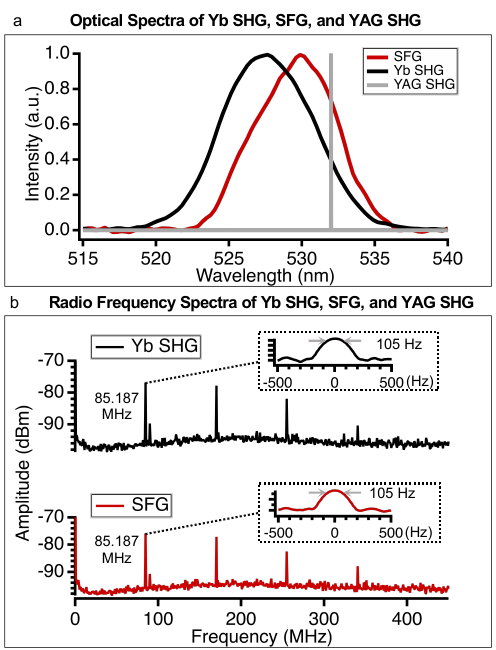}}

\caption{a) Optical spectra of mixed SFG and ultrafast SHG; b) RF spectra of mixed SFG and Yb:fiber SHG. The repetition rate of mixed SFG and Yb:fiber SHG is 85 MHz. 170 MHz, 255 MHz, and 340 MHz are harmonics of the repetition rate. of Yb:fiber SHG and SFG pulses (RBW = 1 MHz). The signal at 90.5 MHz is the local campus radio station. Insets are more detail of the repetition rate of Yb:fiber SHG and SFG signals (measurement RBW = 10 Hz); both widths are therefore limited by the instrument bandwidth.}
\label{fig:fig2}
\end{figure}

Figure \ref{fig:fig2}a presents the optical spectra of the Yb:fiber SHG and SFG beam. The Yb:fiber laser is a frequency comb, so for each comb tooth within the phase-matching bandwidth it will follow Eq.(\ref{eq:1}). Individual comb teeth cannot be resolved with the grating spectrometer. The SFG beam has a center wavelength of 530 nm, while the CW SHG is at 532 nm, and pulsed SHG is at 528 nm. The center wavelength of the SFG beam agrees with the theoretically expected value.  For the ultrafast Yb:fiber SHG process with an interaction length with BBO of 1 mm, the calculated FWHM phase-matching bandwidth\cite{saleh2019fundamentals} is 21.1 nm, while the FWHM of our Yb:fiber source is about 20 nm. With the phase-matching bandwidth slightly larger than the FWHM of the Yb:fiber source, the SHG bandwidth should be 10 nm. For the SFG process, the bandwidth of the Yb:fiber is much greater than that of the CW source, so the predicted SFG FWHM is also 10 nm. The experimentally measured optical FWHM of the SFG signal is 7 nm, and the FWHM of the Yb:fiber SHG signal is 8 nm. The reduced bandwidth of the SHG and SFG processes may be also be due to the spatial walk-off effect by the femtosecond source. 

The walk-off caused by group velocity mismatch is compared with the phase mismatch to see which effect likely dominates the SFG conversion efficiency. The phase mismatch $\Delta k$\ is calculated based on the refractive index provided by SNLO (-3.468/mm) and gives a coherent length $L_{coh}= 2\pi/\Delta k\  = 1.811\,\mathrm{mm}${\cite{saleh2019fundamentals}}.  The group velocity mismatch is $\Delta\beta' = 1/\nu_{SFG} - 1/\nu_{Yb} = 8.66 \times 10^{-11}\,\mathrm{s/m} $. For the $\tau = 97\,\mathrm{fs}$ pulse duration of Yb:fiber laser, the walk-off length caused by group velocity mismatch is therefore $L_{g} = \tau/\Delta\beta' = 1.12\,\mathrm{mm}$. Since $L_{g}<L_{coh}$, the group velocity mismatch will dominate and cause additional conversion efficiency decrease. 

To confirm that the SFG beam is pulsed, a photodetector captured the ultrafast SHG and SFG signal and recorded with the RF spectrum analyzer to get the repetition rate of the beams, shown in Figure \ref{fig:fig2}b. The SFG beam gives an 85 MHz repetition rate, which agrees with the Yb:fiber source. The 3 dB bandwidth for both Yb SHG and SFG signals is 105 Hz, consistent with the lower limit of the spectrum analyzer and confirming that the generated SFG beam has the same pulse repetition rate as the SHG beam.

\begin{figure}[H]
\centering
\fbox{\includegraphics[width=0.8\figurewidthdouble]{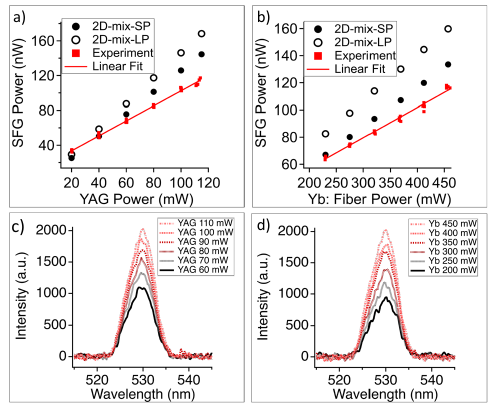}}
\caption{a) SFG power of simulation and experimental result with 470 mW ultrafast Yb:fiber power; b) SFG power of simulation and experimental result when power of YAG (CW) of 110 mW; c) Spectra of SFG beam with a Yb:fiber(ultrafast) power of 470 mW; d) Spectra of SFG beam with CW Nd:YAG power of 110 mW;}
\label{fig:fig3}
\end{figure}

Figure \ref{fig:fig3}a plots the generated SFG power with respect to the CW Nd:YAG laser power while the ultrafast Yb:fiber laser was held at 470 nW and figure \ref{fig:fig3}b plots the generated SFG power while varying the ultrafast Yb:fiber laser power and keeping the CW laser at 110 mW. The data was taken five different times, all plotted on the graph, and least-squares fit to a line (slope and intercept). Figure \ref{fig:fig3}c and \ref{fig:fig3}d show the optical spectra of the SFG light for the same set of powers. As evident from the plot, the spectra of the SFG does not change significantly as the power of the CW laser (c) or ultrafast laser (d) is increased. The SFG power depends linearly on the CW source power and also linearly on the ultrafast source power, which is consistent with an overall second-order process that is first order with each laser source. The peak efficiency for the SFG beam is $\eta_{SFG/CW} = P_{SFG}/P_{CW} = 1.05 \times 10^{-6} $  with respect to CW source power; $\eta_{SFG/pulsed} = P_{SFG}/P_{pulsed} = 2.56 \times 10^{-7}$ with respect to the ultrafast source power.

To complement the experimental results, we performed simulations using SNLO\cite{snlo} to predict conversion efficiencies. The standard procedures available in SNLO do not address the scenario of a pulsed laser, specifically an ultrafast laser, mixing with a CW laser. We therefore compared results using the both the long pulse (2D-mix-LP) and short pulse  (2D-mix-SP) programs with the experimental results. The two-dimensional long-pulse mixing (2D-mix-LP) is a model for nonlinear single-pass mixing for long pulse durations up to CW lasers. It includes the effects of Gaussian spatial profiles, birefringent walk-off, and diffraction but ignores group velocity effects. In 2D-mix-LP, the parameters for SNLO simulation is based on experimental beam parameters and Table \ref{tab: SNLO parameters}.

The two-dimensional short-pulse mixing (2D-mix-SP) is a model for ps and fs pulses that incorporates group velocity effects. It also includes dispersion and diffraction for Gaussian beams. The upper limit for pulse duration in 2D-mix-SP is 100 ps, so the CW laser ``pulse duration'' is set to the upper limit of 100 ps to use this model. The fitting parameters in Table \ref{tab: SNLO parameters} are retrieved from SNLO under the designated phase matching condition. Other input parameters including pulse duration, power, and Gaussian beam parameters are experimentally measured or calculated using the ABCD matrix method. Since the peak power of a Gaussian pulse is about 0.94 times the pulse energy divided by pulse duration, we calculated the simulation's pulse energy based on the CW power and use it to match experimental condition.

\begin{table}[H]
\centering
\caption{\bf Parameters for SNLO Simulation}
\label{tab: SNLO parameters}
\begin{tabular}{c|c|c|c}
\hline
\textbf{Parameter} & CW & Pulsed & SFG \\
\hline
Wavelength (nm) & 1064(o) & 1055(o) & 529.7(e) \\
Phase Velocity & \(c/1.654241\) & \( c/1.654408\) & \(c/1.654364\) \\
Group Velocity &\(c/1.67390\)&\( c/1.67410 \)& \(c/1.70008 \) \\
GDD (fs\(^2\)/mm) & 44.014 & 44.929 & 129.094 \\
Walkoff Angle (mrad) & 0 & 0 & 55.9 \\
\hline
\(d_{\text{eff}}\) (pm/V) & \multicolumn{3}{c}{1.994} \\
\(\Delta k (1/mm) \) & \multicolumn{3}{c}{-3.468} \\
\hline
\end{tabular}
\end{table}

The results given by the 2D-mix-LP are slightly larger in SFG power compared to the 2D-mix-SP in Figures \ref{fig:fig3}a) and \ref{fig:fig3}b). The lack of group velocity mismatch taken into account by the model likely accounts for this, as discussed earlier. Therefore the long-pulse simulated SFG power is expected to be larger than the short-pulse simulated power. However, since the interactive length for SFG is only 1 mm, the group velocity mismatch error is only a relatively small difference between the two simulation types as seen in Figure \ref{fig:fig3}. 

The experimental results are systematically lower in conversion efficiency compared to the 2D-mix-SP simulation. This is likely a result of the non-co-linear geometry used in the experiment as compared to the co-linear geometry of the simulation. In a non-co-linear geometry there is an additional phase mismatch resulting from the finite crossing angle of the beams as they focus on the crystal. In additon, although SNLO takes beam focusing into account, there is some small error in the experimental measurement and subsequent ABCD matrix calculation of beam parameters. According to our ABCD matrix calculation and experimental beam diameter measurement\cite{razorblademethod}, the Rayleigh range $z_R$ of the source is 0.59 mm, which is shorter than the interaction length of 1 mm. The optimal focus and crystal position was found by experimentally optimizing the SFG output power, which leaves a small uncertainty in the beam parameters to put into the SNLO simulation. Therefore we used the ultrafast Yb:fiber SHG experimental result, which the simulation is designed to work for, and slightly adjusted the beam diameter and radius of curvature to get the SHG data to match the simulation output. We then used those adjusted parameters as the beam input parameters in the SFG simulation.

\section{Conclusion and Future Perspectives}

This experiment investigated sum-frequency generation between a CW laser and an ultrafast frequency comb laser. The resulting SFG beam has the repetition rate of the ultrafast laser and the corresponding broad spectra. The total output powers achieved with these input power are low, but sufficient for a cavity-enhanced absorption experiment and other experiments that do not require high power. The SNLO simulation of SFG power agrees with our observation, which will be used to optimize the conversion efficiency in upcoming research. This experiment provides the foundation to use this concept of mixed ultrafast-CW nonlinear frequency generation. By utilizing a commercial tunable CW laser or sets of CW lasers, such as Quantum-Cascade Lasers, External-Cavity Diode Lasers (ECDL), or a CW fiber lasers, paired with appropriate amplifiers if needed, one could achieve frequency tunability by only changing the phase matching angle of BBO crystal in a three-wave mixing process, without implementing cavities or delay stages. The successful realization of SFG suggests that its inverse process, DFG, could also be achieved. With DFG, tunable mid-IR frequencies are accessible. A visible or NIR supercontinuum frequency comb{\cite{ruehl2011ultrabroadband}} could be used as the ultrafast source to attain a mid-IR supercontinuum frequency comb with this setup. For example, with 15 W from a commercially available 1550 nm, narrow-linewidth CW Er-doped fiber laser with amplifier combined with 9 W of our Yb:fiber ultrafast laser at 1060 nm, DFG in a stock, 2 cm long KTP crystal gives 335 mW of ultrafast light at 3.3 $\mu m$. At these wavelengths the walk-off in the KTP crystal is suitable small to allow for use of a longer crystal for higher power. This is sufficient power for many absorption experiments requiring either ultrafast or frequency combs or both.

With the high powers now easily achievable in ultrafast Yb:fiber lasers\cite{li2016high}, ample power for the ultrafast and CW lasers is available to compensate for the lower conversion efficiency and lack of cavity of OPO. Our work has demonstrated the viability of three-wave mixing between an frequency comb and a CW laser, providing support for the core idea and design criteria behind the simplified designs of tunable ultrafast frequency comb lasers.

\begin{acknowledgement}

This work was supported by the U.S. Department of Energy, Office of Science, Office of Basic Energy Sciences, Gas Phase Chemical Physics Program under Award Number DE-SC-0020268. We also thank Walker Jones, Todd Eliason and Uyen M. Ta for their help.

\end{acknowledgement}

\bibliography{sfgpaper}

\end{document}